\documentclass[12pt,draftcls,onecolumn]{IEEEtran}

\usepackage[dvips]{color}
\usepackage{graphicx}
\usepackage{amsmath}
\begin{document}
%
\title{Improving Physical-Layer Security in Wireless Communications Using Diversity Techniques}
%
%
%

\author{Yulong~Zou,~\IEEEmembership{Senior Member,~IEEE,}
Jia Zhu,
Xianbin Wang,~\IEEEmembership{Senior Member,~IEEE,}
and Victor C.M. Leung,~\IEEEmembership{Fellow,~IEEE}

\thanks{Y. Zou and J. Zhu are with the Nanjing University of Posts and Telecomm., China. (Email: \{yulong.zou, jiazhu\}@njupt.edu.cn)}
\thanks{X. Wang is with the University of Western Ontario, Canada. (Email: xianbin.wang@uwo.ca)}
\thanks{V. C.M. Leung is with the University of British Columbia, Canada. (Email: vleung@ece.ubc.ca)}

}

\maketitle

\begin{abstract}
Due to the broadcast nature of radio propagation, the wireless transmission can be readily overheard by unauthorized users for interception purposes and is thus highly vulnerable to eavesdropping attacks. To this end, physical-layer security is emerging as a promising paradigm to protect the wireless communications against eavesdropping attacks by exploiting the physical characteristics of wireless channels. This article is focused on the investigation of diversity techniques to improve the physical-layer security, differing from the conventional artificial noise generation and beamforming techniques which typically consume additional power for generating artificial noise and exhibit high implementation complexity for beamformer design. We present several diversity approaches to improve the wireless physical-layer security, including the multiple-input multiple-output (MIMO), multiuser diversity, and cooperative diversity. To illustrate the security improvement through diversity, we propose a case study of exploiting cooperative relays to assist the signal transmission from source to destination while defending against eavesdropping attacks. We evaluate the security performance of cooperative relay transmission in Rayleigh fading environments in terms of secrecy capacity and intercept probability. It is shown that as the number of relays increases, the secrecy capacity and intercept probability of the cooperative relay transmission both improve significantly, implying the advantage of exploiting cooperative diversity to improve the physical-layer security against eavesdropping attacks.

\end{abstract}


\IEEEpeerreviewmaketitle

\section{Introduction}

\IEEEPARstart {I}{n} wireless networks, the transmission between legitimate users can be easily overheard by an eavesdropper for interception due to the broadcast nature of wireless medium, making the wireless transmission highly vulnerable to eavesdropping attacks. In order to achieve the confidential transmission, existing communications systems typically adopt the cryptographic techniques to prevent an eavesdropper from tapping the data transmission between legitimate users [1], [2]. By considering the symmetric key encryption as an example, the original data (called plaintext) is first encrypted at source node by using an encryption algorithm along with a secret key that is shared with destination node only. Then, the encrypted plaintext (also known as ciphertext) is transmitted to destination that will decrypt its received ciphertext with the pre-shared secret key. In this way, even if an eavesdropper overhears the ciphertext transmission, it is still difficult to interpret the plaintext by the eavesdropper from its intercepted ciphertext without the secret key. It is pointed out that the ciphertext transmission is not perfectly secure, since the ciphertext can still be decrypted by an eavesdropper with the exhaustive key search, which is also known as the brute-force attack. To this end, physical-layer security is emerging as an alternative paradigm to protect the wireless communications against eavesdropping attacks, including the brute-force attack.

Physical-layer security work was pioneered by Wyner in [3], where a discrete memoryless wiretap channel was examined for secure
communications in the presence of an eavesdropper. It was proved in [3] that the perfectly secure data transmission can be achieved if the channel capacity of the main link (from source to destination) is higher than that of the wiretap link (from source to eavesdropper). Later on, in [4], the Wyner's results were extended from the discrete memoryless wiretap channel to the Gaussian wiretap channel, where a so-called \emph{secrecy capacity} was developed and shown as the difference between the channel capacity of the main link and that of the wiretap link. If the secrecy capacity falls below zero, the transmission from source to destination becomes insecure and the eavesdropper would succeed in intercepting the source transmission, i.e., an intercept event occurs. In order to improve the transmission security against eavesdropping attacks, it is of importance to reduce the probability of occurrence of an intercept event (called \emph{intercept probability}) through enlarging the secrecy capacity. However, in wireless communications, the secrecy capacity severely degrades due to the fading effect.

{As a consequence, there are extensive works aimed at increasing the secrecy capacity of wireless communications by exploiting the multiple antennas [5] and cooperative relays [6]. Specifically, the multiple-input multiple-output (MIMO) wiretap channel was studied in [7] to enhance the wireless secrecy capacity in fading environments. In [8], the cooperative relays were examined for improving the physical-layer security in terms of the secrecy rate performance. A hybrid cooperative beamforming and jamming approach was investigated in [9] to enhance the wireless secrecy capacity, where partial relay nodes are allowed to assist the source transmission to the legitimate destination with the aid of distributed beamforming, while the remaining relay nodes are used to transmit artificial noise for confusing the eavesdropper. More recently, a joint physical-application layer security framework was proposed in [10] for improving the security of wireless multimedia delivery by simultaneously exploiting the physical-layer signal processing techniques as well as the upper-layer authentication and watermarking methods. In [11], the error-control coding for secrecy was discussed for achieving the physical-layer security. Additionally, in [12] and [13], the physical-layer security was further investigated in emerging cognitive radio networks.}

At the time of writing, most research efforts are devoted to examining the artificial noise and beamforming techniques to combat the eavesdropping attacks, which, however, consume additional power resources for generating artificial noise and increase the computational complexity for performing beamformer design. Therefore, this paper is motivated to enhance the physical-layer security through diversity techniques without additional power costs, including the MIMO, multiuser diversity, and cooperative diversity, aiming at increasing the capacity of the main channel while degrading the wiretap channel. For illustration purposes, we present a case study of exploiting cooperative relays to improve the physical-layer security against eavesdropping attacks, where the best relay is selected and used to participate in forwarding the signal transmission from source to destination. We evaluate the secrecy capacity and intercept probability of proposed cooperative relay transmission in Rayleigh fading environments. It is shown that upon increasing number of relays, the security performance of cooperative relay transmission significantly improves in terms of the secrecy capacity and intercept probability. This confirms the advantage of using cooperative relays to protect the wireless communications against eavesdropping attacks.

The remainder of this article is organized as follows. Section II presents the system model of physical-layer security in wireless communications. Next, in Section III, we are focused on the physical-layer security enhancement through diversity techniques, including the MIMO, multiuser diversity, and cooperative diversity. For the purpose of illustrating the security improvement through diversity, Section IV presents a case study of exploiting cooperative relays to assist the signal transmission from source to destination against eavesdropping attacks. Finally, we provide some concluding remarks in Section V.

\section{Physical-Layer Security in Wireless Communications}
\begin{figure}
  \centering
  {\includegraphics[scale=0.85]{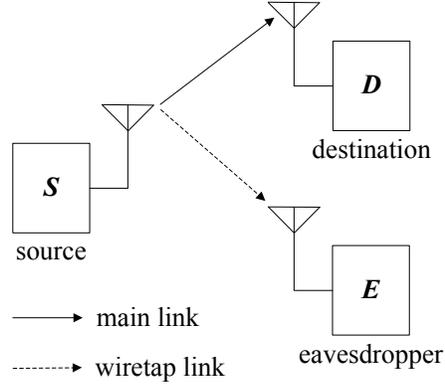}\\
  \caption{A wireless communications scenario consisting of one source and one destination
  in the presence of an eavesdropping attack.}\label{Fig1}}
\end{figure}
Fig. 1 shows a wireless communications scenario with one source and
one destination in the presence of an eavesdropper, where the solid
and dash lines represent the main channel (from source to
destination) and the wiretap channel (from source to eavesdropper),
respectively. When the source node transmits its signal to
destination, an eavesdropper may overhear such transmission due to
the broadcast nature of wireless medium. Considering the fact that
today's wireless systems are highly standardized, the eavesdropper
can readily obtain the transmission parameters, including the signal
waveform, coding and modulation scheme, encryption algorithm, and so
on. Also, the secret key may be figured out at the eavesdropper e.g.
through the exhaustive search. Thus, the source signal could be
interpreted at the eavesdropper by decoding its overheard signal,
leading the legitimate transmission to be insecure.

As a result, physical-layer security emerges as an alternative means
to achieve the perfect transmission secrecy from source to
destination. In the physical-layer security literature [3] and [4], a so-called ``secrecy capacity" is developed and shown as the difference between the capacities of main link and wiretap link. It
has been proven that perfect secrecy is achieved if the secrecy
capacity is positive, meaning that when the main channel capacity is
larger than the wiretap channel capacity, the transmission from
source to destination can be perfectly secure. This can be explained
by using the Shannon coding theorem from which a receiver is
impossible to recover the source signal if the channel capacity
(from source to the receiver) is smaller than the data rate. Thus,
given a positive secrecy capacity, the data rate can be adjusted
between the capacities of main and wiretap channels so that the
destination node successfully decodes the source signal and the
eavesdropper fails to decode. However, if the secrecy capacity is
negative (i.e., the main channel capacity falls below the wiretap
channel capacity), the eavesdropper is more likely than the
destination to succeed in decoding the source signal. In an
information-theoretic sense, when the main channel capacity becomes
smaller than the wiretap channel capacity, it is impossible to
guarantee that the destination succeeds and the eavesdropper fails
to decode the source signal. Therefore, an intercept event is viewed
to occur when the secrecy capacity falls below zero, and the
probability of occurrence of intercept event is called intercept
probability throughout this article.

At present, most existing work is focused on improving the
physical-layer security by generating artificial noise to confuse
the eavesdropping attack, where the artificial noise is
sophisticatedly produced such that only the eavesdropper is
interfered and the desired destination can easily cancel out such
noise without performance degradation. More specifically, given a
main channel matrix ${\textbf{\textrm{H}}_m}$, the artificial noise
(denoted by ${\textbf{\textrm{w}}}_n$) is designed in the null space of matrix ${\textbf{\textrm{H}}_m}$ such that
${\textbf{\textrm{H}}_m}{\textbf{\textrm{w}}}_n=0$, making the
desired destination unaffected by the noise. Since the wiretap
channel is independent of the main channel, the null space of
wiretap channel is in general different from that of main channel
and thus the eavesdropper can not null out the artificial noise,
which results in the performance degradation at the eavesdropper.
Notice that the above-mentioned null space based noise generation
approach needs the knowledge of main channel
${\textbf{\textrm{H}}_m}$ only, which can be further optimized if
the wiretap channel information is also known. It needs to be
pointed out that additional power resources are required for
generating artificial noise to confuse the eavesdropper. For a fair
comparison, the total transmit power of artificial noise and desired
signal should be constrained. Also, the power allocation between the
artificial noise and desired signal is important and should be
adapted to the main and wiretap channels to optimize the
physical-layer security performance e.g. in terms of secrecy
capacity. Differing from the artificial noise generation approach,
this article is mainly focused on the investigation of diversity
techniques for enhancing the physical-layer security.

\section{Diversity for Physical-Layer Security}
In this section, we present several diversity techniques to improve
the physical-layer security against eavesdropping attacks.
Traditionally, diversity techniques are exploited to increase the
transmission reliability, which also have great potential to enhance
the wireless security. In the following, we will discuss the
physical-layer security improvement through the use of MIMO,
multiuser diversity, and cooperative diversity, respectively. {Notice that the MIMO and multiuser diversity mechanisms are generally applicable to various cellular and Wi-Fi networks, since the cellular and Wi-Fi networks typically consist of multiple users and, moreover, today's cellular and Wi-Fi devices are equipped with multiple antennas. In contrast, the cooperative diversity mechanism is only applicable to some advanced cellular and Wi-Fi networks that have adopted the relay architecture e.g. the long term evolution (LTE)-advanced system and IEEE 802.16j/m, where relay stations are introduced to assist the wireless data transmission.}

\subsection{MIMO Diversity}
\begin{figure}
  \centering
  {\includegraphics[scale=0.8]{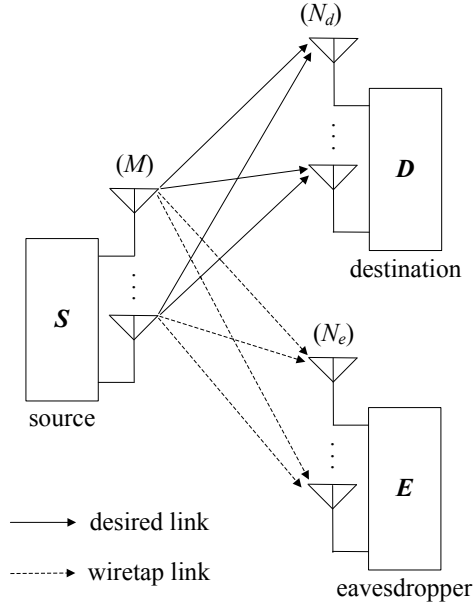}\\
  \caption{A multiple-input multiple-output (MIMO) wireless system consisting of one source
  and one destination in the presence of an eavesdropping attack.}\label{Fig2}}
\end{figure}

This subsection presents MIMO diversity for physical-layer security
of wireless transmission against eavesdropping attacks. As shown in
Fig. 2, all the network nodes are equipped with multiple antennas,
where $M$, $N_d$ and $N_e$ represent the number of antennas at
source, destination and eavesdropper, respectively. As is known,
MIMO has been shown as an effective means to combat wireless fading
and increase the capacity of wireless channel. However, the
eavesdropper can also exploit the MIMO structure to enlarge the
capacity of wiretap channel from source to eavesdropper. Thus,
without a proper design, it may fail to increase the secrecy
capacity of wireless transmission with MIMO. For example, if the
conventional open-loop space-time block coding is considered, the
destination should first estimate the main channel matrix
${\textbf{\textrm{H}}_m}$ and then perform the space-time decoding
process with an estimated ${\hat {\textbf{\textrm{H}}}}_m$, leading
the diversity gain to be achieved for the main channel. Similarly,
the eavesdropper can also estimate the wiretap channel matrix
${\textbf{\textrm{H}}_w}$ and then conduct the corresponding
space-time decoding algorithm to obtain diversity gain for the
wiretap channel. Hence, the conventional space-time block coding is
not effective to improve the physical-layer security against
eavesdropping attacks.

Generally speaking, if the source node transmits its signal to the
desired destination with $M$ antennas, the eavesdropper will also
receive $M$ signal copies for interception purposes. In order to
defend against eavesdropping attacks, the source node should adopt a
preprocess that needs to be adapted to the main and wiretap channels
${\textbf{\textrm{H}}_m}$ and ${\textbf{\textrm{H}}_w}$ such that
the diversity gain can be achieved at destination only whereas the
eavesdropper benefits nothing from the multiple transmit antennas at
source. This means that an \emph{adaptive transmit process} should
be included at the source node to increase the main channel capacity
while decreasing the wiretap channel capacity. Ideally, the
objective of such adaptive transmit process is to maximize the
secrecy capacity of MIMO transmission, which, however, requires the
channel state information (CSI) of both main and wiretap links
(i.e., ${\textbf{\textrm{H}}_m}$ and ${\textbf{\textrm{H}}_w}$). In
practice, the wiretap channel information ${\textbf{\textrm{H}}_w}$
may be unavailable, since the eavesdropper is usually passive and
keeps silent. If only the main channel information
${\textbf{\textrm{H}}_m}$ is known, the adaptive transmit process
can be designed to maximize the main channel capacity, which does
not require the knowledge of wiretap channel
${\textbf{\textrm{H}}_w}$. Since the adaptive transmit process is
optimized based on the main channel information
${\textbf{\textrm{H}}_m}$ and the wiretap channel is typically
independent of the main channel, the main channel capacity will be
significantly increased with MIMO and no improvement will be
achieved for the wiretap channel capacity.

As for the aforementioned adaptive transmit process, we here present
three main concrete approaches: transmit beamforming, power
allocation, and transmit antenna selection. The transmit beamforming
is a signal processing technique by combining multiple transmit
antennas at the source node in such a way that desired signals
transmit in a particular direction to destination. Considering that
the eavesdropper and destination generally lie in different
directions relative to the source node, the desired signals (with
transmit beamforming) that are received at the eavesdropper will
experience destructive interference and become very weak. Thus, the
transmit beamforming is effective to defend against eavesdropping attacks when the destination and eavesdropper are spatially
separated. The power allocation is to maximize the main
channel capacity (or secrecy capacity if both
${\textbf{\textrm{H}}_m}$ and ${\textbf{\textrm{H}}_w}$ are known)
by allocating the transmit power among $M$ antennas at source. In
this way, the secrecy capacity of MIMO transmission will be
significantly increased, showing the security benefits of using
power allocation against eavesdropping attacks. In addition, the
transmit antenna selection is also able to improve the
physical-layer security of MIMO wireless systems. Depending on
whether the global CSI of main and wiretap channels (i.e.,
${\textbf{\textrm{H}}_m}$ and ${\textbf{\textrm{H}}_w}$) is
available, an optimal transmit antenna at the source node will be
selected and used to transmit source signals. More specifically, if
both ${\textbf{\textrm{H}}_m}$ and ${\textbf{\textrm{H}}_w}$ are
available, a transmit antenna with the highest secrecy capacity will
be chosen. Studying the case of the global CSI available will provide a theoretical upper bound on the security performance of wireless systems. Notice that the CSI of wiretap channels may be estimated and obtained by monitoring the eavesdroppers' transmissions as discussed in [8] and [14]. If only ${\textbf{\textrm{H}}_m}$ is known, the transmit antenna selection is to maximize the main channel capacity. One can observe that the above-mentioned three approaches (i.e., transmit beamforming, power allocation, and transmit antenna
selection) all have great potential to improve the physical-layer
security of MIMO wireless systems against eavesdropping attacks.

\subsection{Multiuser Diversity}
\begin{figure}
  \centering
  {\includegraphics[scale=0.85]{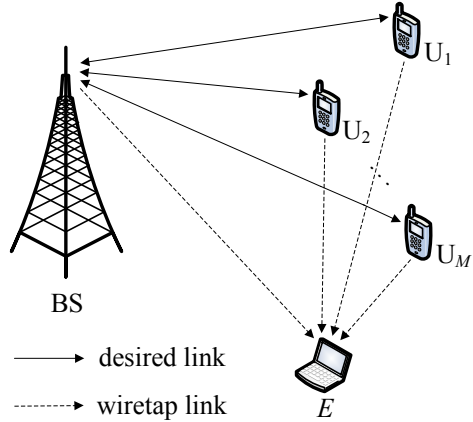}\\
  \caption{A multiuser wireless communications system consisting of one base station (BS)
  and multiple users in the presence of an eavesdropper.}\label{Fig3}}
\end{figure}

This subsection discusses the multiuser diversity for improving
physical-layer security. Fig. 3 shows that a base station (BS)
serves multiple users where $M$ users are denoted by
${\cal{U}}=\{U_i|i=1,2,\cdots,M\}$. In cellular networks, $M$ users
typically communicate with BS with an orthogonal multiple access
mechanism such as the orthogonal frequency division multiple access
(OFDMA) and time division multiple access (TDMA). Taking the OFDMA
as an example, the OFDM subcarriers are allocated to different
users. In other words, given an OFDM subcarrier, we need to
determine which user should be assigned to access and use the
subcarrier for data transmission. Traditionally, a user with the
highest throughput is selected to access the given OFDM subcarrier,
aiming at maximizing the transmission capacity. This relies on the
knowledge of main channel information ${\textbf{\textrm{H}}_m}$ only
and can provide the significant multiuser diversity gain for
performance improvement. However, if a user is far away from BS and
experiences severe propagation loss and deep fading, it may have no
chance to be selected as the ``best" user for channel access. To
this end, user fairness should be further considered in the
multiuser scheduling, where two competing interests need to be
balanced: maximizing the main channel capacity while at the same
time guaranteeing each user with certain opportunities to access the
channel.

With the multiuser scheduling, a user is first selected to access a
channel (i.e., an OFDM subcarrier in OFDMA or a time slot in TDMA)
and then starts transmitting its signal to BS. Meanwhile, due to the
broadcast nature of wireless transmission, the eavesdropper
overhears such transmission and attempts to interpret the source
signal. In order to effectively defend against the eavesdropping
attack, the multiuser scheduling should be performed to minimize the
wiretap channel capacity while maximizing the main channel capacity,
which requires the CSI of both main and wiretap links. If only the
main channel information ${\textbf{\textrm{H}}_m}$ is available, we
may consider the use of conventional multiuser scheduling where the
wiretap channel information {${\textbf{\textrm{H}}_w}$} is not taken
into account. It needs to be pointed out the conventional multiuser
scheduling still has great potential to enhance the physical-layer
security, since the main channel capacity is significantly improved
with conventional multiuser scheduling while the wiretap channel
capacity remains the same.

\subsection{Cooperative Diversity}
\begin{figure}
  \centering
  {\includegraphics[scale=0.85]{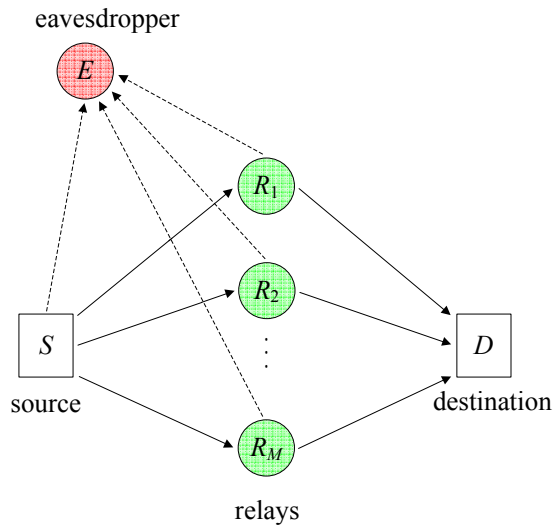}\\
  \caption{A cooperative diversity system consisting of one source, $M$ relays, and
  one destination in the presence of an eavesdropper.}\label{Fig4}}
\end{figure}

In this subsection, we are mainly focused on the cooperative
diversity for wireless security against eavesdropping attacks. Fig. 4
shows a cooperative wireless network including one source, $M$
relays, and one destination in the presence of an eavesdropper,
where $M$ relays are exploited to assist the signal transmission
from source to destination. To be specific, the source node first
transmits its signal to $M$ relays that then forward their received
source signals to destination. At present, there are two basic relay
protocols: amplify-and-forward (AF) and decode-and-forward (DF). In
the AF protocol, a relay node simply amplifies and retransmits its
received noisy version of the source signal to the destination. In
contrast, the DF protocol requires the relay node to decode its
received signal and forward its decoded outcome to the destination
node. It is concluded that the multiple relays assisted source
signal transmission consists of two steps: 1) the source node
broadcasts its signal, and 2) relay nodes retransmit their received
signals. Each of the two transmission steps is vulnerable to
eavesdropping attack and needs to be carefully designed to prevent
an eavesdropper from intercepting the source signal.

Typically, the main channel capacity with multiple relays can be
significantly increased by using cooperative beamforming. More
specifically, multiple relays can form a virtual antenna array and
cooperate with each other to perform transmit beamforming such that
the signals received at the intended destination experience
constructive interference while the others (received at
eavesdropper) experience destructive interference. One can observe
that with the cooperative beamforming, the received signal strength
of destination will be much higher than that of eavesdropper,
implying the physical-layer security improvement. In addition to the
aforementioned cooperative beamforming, the best relay selection is
another approach to improve the wireless transmission security
against eavesdropping attacks. In the best relay selection, a relay
node with the highest secrecy capacity (or highest main channel
capacity if only the main channel information is available) is
chosen to participate in assisting the signal transmission from
source to destination. In this way, the cooperative diversity gain
will be achieved for the physical-layer security enhancement.

\section{Case Study: Security Evaluation of Cooperative Relay Transmission}
In this section, we present a case study to show the physical-layer
security improvement by exploiting cooperative relays, {where only a
single best relay will be selected to assist the signal transmission
from source to destination. This differs from existing research
efforts in [8], where multiple cooperative relays participate in
forwarding the source signal to destination.} For the comparison
purpose, we first consider the conventional direct transmission as a
benchmark scheme, where the source node directly transmits its
signal to destination without relay. Meanwhile, an eavesdropper is
present and attempts to intercept the signal transmission from
source to destination. As discussed in [3] and [4], the secrecy
capacity of conventional direct transmission is shown as the
difference between the capacities of main channel (from source to
destination) and wiretap channel (from source to eavesdropper),
which is written as
\begin{equation}\label{euqa1} {C_s} = {\log _2}\left( {1 +
\frac{{P|{h_{sd}}{|^2}}}{{{N_0}}}} \right) - {\log _2}\left( {1 +
\frac{{P|{h_{se}}{|^2}}}{{{N_0}}}} \right),
\end{equation}
where $P$ is the transmit power at source, $N_0$ is the variance of
additive white Gaussian noise (AWGN), $\gamma_s=P/N_0$ is
regarded to as the signal-to-noise ratio (SNR), and $h_{sd}$ and $h_{se}$ represent fading coefficients of the channel from source to
destination and that from source to eavesdropper, respectively.
Presently, there are three commonly used fading models (i.e.,
Rayleigh, Rician and Nakagami) and we consider the use of Rayleigh
fading model to characterize the main and wiretap channels. Thus,
$|h_{sd}{|^2}$ and $|h_{se}{|^2}$ are independent and exponentially
distributed random variables with means $\sigma _{sd}^2$ and
$\sigma _{se}^2$, respectively. Also, an ergodic secrecy capacity of
the direct transmission can be obtained by averaging the
instantaneous secrecy capacity ${C_s}^{+}$ over the fading
coefficients $h_{sd}$ and $h_{se}$, where $C_s^ +  = \max \left(
{{C_s},0} \right)$. In addition, if the secrecy capacity $C_s$ falls
below zero, the source transmission becomes insecure and the
eavesdropper will succeed in intercepting the source signal. {Thus, using Eq. (1) and denoting $x = |{h_{sd}}{|^2}$ and $y = |{h_{se}}{|^2}$}, an intercept probability of the direct transmission can be given by
\begin{equation}\label{equa2}
\begin{split}
{P_{{\textrm{intercept}}}} &= \Pr\left( {{C_s} < 0} \right) \\
&=\Pr \left( {|{h_{sd}}{|^2} < |{h_{se}}{|^2}} \right) \\
&=\iint\limits_{x < y} {\frac{1}{{\sigma _{sd}^2}{\sigma _{se}^2}}\exp ( - \frac{x}{{\sigma _{sd}^2}} - \frac{y}{{\sigma _{se}^2}})dxdy} \\
&= \frac{{\sigma _{se}^2}}{{\sigma _{sd}^2+ \sigma _{se}^2}},
\end{split}
\end{equation}
where the third equation arises from the fact that random variables $|h_{sd}|^2$ and $|h_{se}|^2$ are independent exponentially distributed, and $\sigma^2_{sd}$ and $\sigma^2_{se}$ are the expected values of $|h_{sd}{|^2}$ and $|h_{se}{|^2}$, respectively. As can be observed from Eq. (2), the intercept probability of
conventional direct transmission is independent of the transmit
power $P$, meaning that increasing the transmit power cannot improve
the physical-layer security in terms of intercept probability. This
motivates us to explore the use of cooperative relays to decrease
the intercept probability. For notational convenience, let
$\lambda_{me}$ represent the ratio of {average main channel gain}
$\sigma^2_{sd}$ to eavesdropper's {average channel gain} $\sigma^2_{se}$,
i.e., $\lambda_{me}=\sigma^2_{sd}/\sigma^2_{se}$, which is referred
to as the main-to-eavesdropper ratio (MER) throughout this article.
In the following, we present the cooperative relay transmission
scheme where multiple relays are used to assist the signal
transmission from source to destination. Here, the AF relaying
protocol is considered and only the best relay will be selected to
participate in forwarding the source signal to destination. To be
specific, the source node first broadcasts its signal to $M$ relays.
Then, the best relay node will be chosen to forward a scaled version
of its received signal to destination [15]. Notice that during the above-mentioned cooperative relay transmission process, the total amount of transmit power at source and relay should be constrained to $P$ to make a fair comparison with the conventional direct transmission scheme. We here consider the equal-power allocation and thus the transmit power at source and relay is given by $P/2$.

Now, given $M$ relays, it is crucial to determine which relay should
be selected as the best one to assist the source signal
transmission. Ideally, the best relay selection should aim to
maximize the secrecy capacity, which, however, requires the CSI of
both main and wiretap channels. Since the eavesdropper is passive
and the wiretap channel information is difficult to obtain in
practice, we consider the main channel capacity as the objective of
best relay selection, which relies on the knowledge of main channel
only. Accordingly, the best relay selection criterion with AF protocol is expressed as
\begin{equation}\label{equa3}
{\textrm{Best Relay}}
= \arg \mathop {\max }\limits_{i \in {\cal{R}}} \frac{{|{h_{si}}{|^2}|{h_{id}}{|^2}}}{{|{h_{si}}{|^2}
+ |{h_{id}}{|^2}}},
\end{equation}
where ${\cal{R}}$ denotes a set of $M$ relays, and $|{h_{si}}{|^2}$
and $|{h_{id}}{|^2}$ represent fading coefficients of the channel
from source to relay $R_i$ and that from relay $R_i$ to destination,
respectively. One can see from Eq. (3) that the proposed best relay
selection criterion only requires the main channel information
$|{h_{si}}{|^2}$ and $|{h_{id}}{|^2}$, with which the main channel
capacity is maximized. Since the main and wiretap channels are
independent of each other, the wiretap channel capacity will benefit
nothing from the proposed best relay selection. Similar to Eq. (1),
the secrecy capacity of best relay selection scheme can be obtained
through subtracting the main channel capacity by the corresponding
wiretap channel capacity. Also, the intercept probability of best
relay selection is easily determined by computing the probability
that the secrecy capacity becomes less than zero.
\begin{figure}
  \centering
  {\includegraphics[scale=0.85]{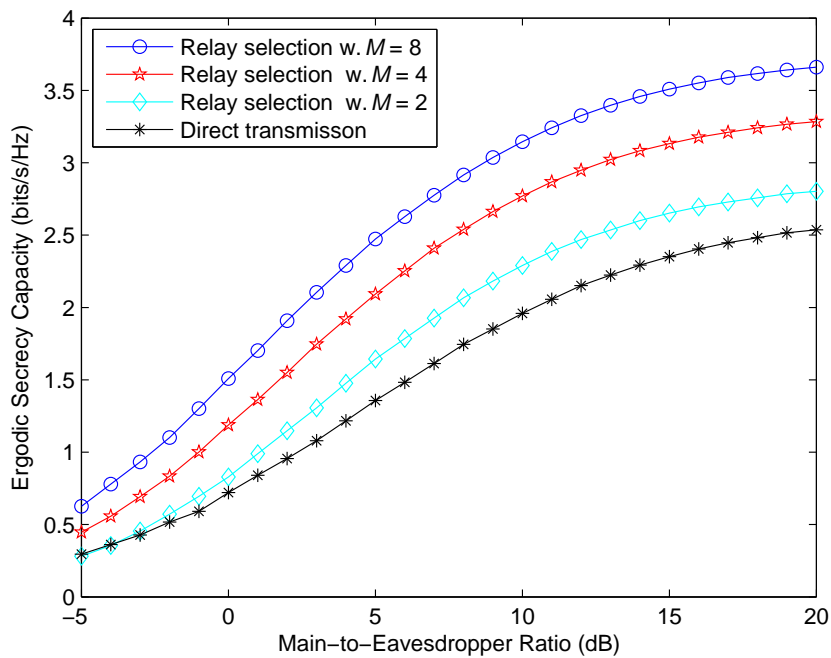}\\
  \caption{Ergodic secrecy capacity versus MER of the direct transmission and best relay
  selection schemes with $\gamma_s=12{\textrm{dB}}$, $\sigma^2_{sd}=0.5$, and
  $\sigma^2_{sr}=\sigma^2_{rd}=2$.}\label{Fig5}}
\end{figure}

In Fig. 5, we provide the ergodic secrecy capacity comparison
between the conventional direct transmission and proposed best relay
selection schemes for different number of relays $M$ with
$\gamma_s=12{\textrm{dB}}$, $\sigma^2_{sd}=0.5$, and
$\sigma^2_{sr}=\sigma^2_{rd}=2$. It is shown from Fig. 5 that for
the cases of $M=2$, $M=4$ and $M=8$, the ergodic secrecy capacity of best relay selection scheme is always higher than that of direct
transmission, showing the wireless security benefits of using
cooperative relays. Also, as the number of relays $M$ increases from
$M=2$ to $M=8$, the ergodic secrecy capacity of best relay selection
scheme significantly increases. This means that increasing the
number of cooperative relays can improve the physical-layer security
of wireless transmission against eavesdropping attacks.

\begin{figure}
  \centering
  {\includegraphics[scale=0.85]{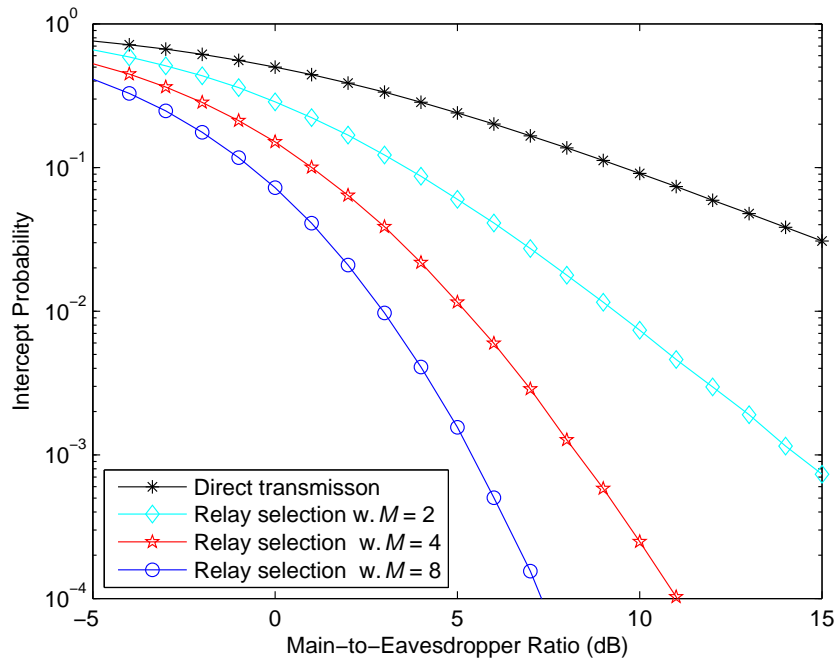}\\
  \caption{Intercept probability versus MER of the direct transmission and best relay selection schemes with $\gamma_s=12{\textrm{dB}}$, $\sigma^2_{sd}=0.5$, and $\sigma^2_{sr}=\sigma^2_{rd}=2$.}\label{Fig6}}
\end{figure}

Fig. 6 shows the intercept probability versus MER of the conventional direct transmission and proposed best relay selection schemes for
different number of relays $M$ with $\gamma_s=12{\textrm{dB}}$, $\sigma^2_{sd}=0.5$, and $\sigma^2_{sr}=\sigma^2_{rd}=2$. Notice that the intercept probability is obtained by calculating the rate of occurrence of an intercept event that the capacity of the main channel falls below that of the wiretap channel. Observe from Fig. 6 that the best relay selection scheme outperforms the conventional direct transmission in terms of intercept probability. Moreover, as the number of cooperative relays $M$ increases from $M=2$ to $M=8$, the intercept probability improvement of best relay selection over direct transmission becomes much more significant. It is also shown from Fig. 6 that the slope of the intercept probability curve of best relay selection scheme in high MER regions becomes steeper with an increasing number of relays. In other words, as the number of relays increases, the intercept probability of best relay selection scheme decreases at much higher speed with an increasing MER. This further confirms that the diversity gain is achieved by the proposed relay selection scheme for the physical-layer security improvement.

\section{Conclusion}
This article studied the physical-layer security of wireless communications and presented several diversity techniques for improving the wireless security against eavesdropping attacks. We discussed the use of multiple-input multiple-output (MIMO), multiuser diversity, and cooperative diversity for the sake of increasing the secrecy capacity of wireless transmission. To illustrate the security benefits through diversity, we proposed a case study of the physical-layer security in cooperative wireless networks with multiple relays, where the best relay was selected to participate in forwarding the signal transmission from source to destination. The secrecy capacity and intercept probability of the conventional direct transmission and the proposed best relay selection schemes were evaluated in Rayleigh fading environments. It was shown that the best relay selection scheme outperforms the direct transmission in terms of both the secrecy capacity and intercept probability. Moreover, as the number of cooperative relays increases, the security improvement of the best relay selection scheme over direct transmission becomes much more significant.

{Although extensive research efforts have been devoted to the wireless physical-layer security, many challenging but interesting issues still remain open for future work. Specifically, most of the existing works in this subject are focused on enhancing the wireless secrecy capacity against the eavesdropping attack only, but have neglected the joint consideration of different types of the wireless physical-layer attacks, including both the eavesdropping and denial-of-service (DoS) attacks. It will be of high importance to explore new techniques of jointly defending against multiple different wireless attacks. Furthermore, the security, reliability and throughput are the main driving factors for the research and development of next-generation wireless networks, which are typically coupled and affect each other. For example, the security of wireless physical layer may be improved by generating the artificial noise for confusing the eavesdropping attack, which, however, comes at the expense of degrading the wireless reliability and throughput performance, since the artificial noise generation consumes some power resources and less transmit power becomes available for the desired information transmission. Thus, it will be of interest to investigate the joint optimization of security, reliability and throughput for the wireless physical layer, which is a challenging issue to be solved in the future. }

\ifCLASSOPTIONcaptionsoff
  \newpage
\fi

\appendices

\clearpage

\end{document}